\documentclass[9pt,twocolumn,twoside]{optica}
\setboolean{shortarticle}{false}
\setboolean{minireview}{false}
\DeclareUnicodeCharacter{1F6C8}{\info}

\title{Spatio-temporal Vision Transformer for Super-resolution Microscopy}

\author[1,2]{Charles N. Christensen}
\author[1]{Meng Lu}
\author[1]{Edward N. Ward}
\author[2]{Pietro Lio}
\author[1,*]{Clemens F. Kaminski}

\affil[1]{University of Cambridge, Department of Chemical Engineering and Biotechnology, Laser Analytics Group, Cambridge, UK}
\affil[2]{University of Cambridge, Department of Computer Science and Technology, Artificial Intelligence Group, Cambridge, UK}

\affil[*]{Corresponding author: cfk23@cam.ac.uk.  $\quad$ -- $\quad$ Interactive implementation and source code: \url{http://vsr-sim.github.io}}
\usepackage{booktabs}
\usepackage{multirow}
\usepackage{placeins}
\usepackage{pdfpages}

\RequirePackage{xspace}
\makeatletter
\DeclareRobustCommand\onedot{\futurelet\@let@token\@onedot}
\def\@onedot{\ifx\@let@token.\else.\null\fi\xspace}

\def\etal{\emph{et al}\onedot}

\begin{document}

\begin{abstract}

    Structured illumination microscopy (SIM) is an optical super-resolution
    technique that enables live-cell imaging beyond the diffraction limit.
    Reconstruction of SIM data is prone to artefacts, which becomes problematic
    when imaging highly dynamic samples because previous methods rely on the
    assumption that samples are static. We propose a new transformer-based
    reconstruction method, VSR-SIM, that uses shifted 3-dimensional window
    multi-head attention in addition to channel attention mechanism to tackle
    the problem of video super-resolution (VSR) in SIM. The attention mechanisms
    are found to capture motion in sequences without the need for common motion
    estimation techniques such as optical flow.  We take an approach to training
    the network that relies solely on simulated data using videos of natural
    scenery with a model for SIM image formation.  We demonstrate a use
    case enabled by VSR-SIM referred to as rolling SIM imaging, which increases
    temporal resolution in SIM by a factor of 9. Our method can be applied to
    any SIM setup enabling precise recordings of dynamic processes in biomedical
    research with high temporal resolution.

\end{abstract}

\vspace{-100pt}
\maketitle
\section{Introduction}

Optical microscopy is limited by the diffraction of light occurring in the
optics of imaging systems. For visible light, the diffraction limit, also known
as the Abbe resolution limit \cite{mccutchen1967superresolution}, is around 200
nm laterally. Structured illumination microscopy (SIM) is an optical microscopy technique
that can achieve a two-fold spatial resolution improvement, thus enabling
sub-diffraction limit imaging -- a regime important for biomedical imaging
\cite{iqbal2010tau}. Furthermore, SIM is live-cell compatible as it can be
performed at relatively low excitation power. A significant challenge in
applying SIM, however, remains the computational reconstruction of the acquired data into
super-resolved images. The reconstruction problem in SIM is an inverse problem
similar to deconvolution \cite{strohl2015joint} but makes use of shifted high
frequency information. The frequency-shifted signals are obtained by
illuminating the sample with a temporal sequence of illumination patterns,
generally sinusoidal fringes with varying orientations and phase shifts, and an
image is captured for each respective pattern.

\begin{figure}[t!]
	\centering
	\includegraphics[width=0.5\textwidth]{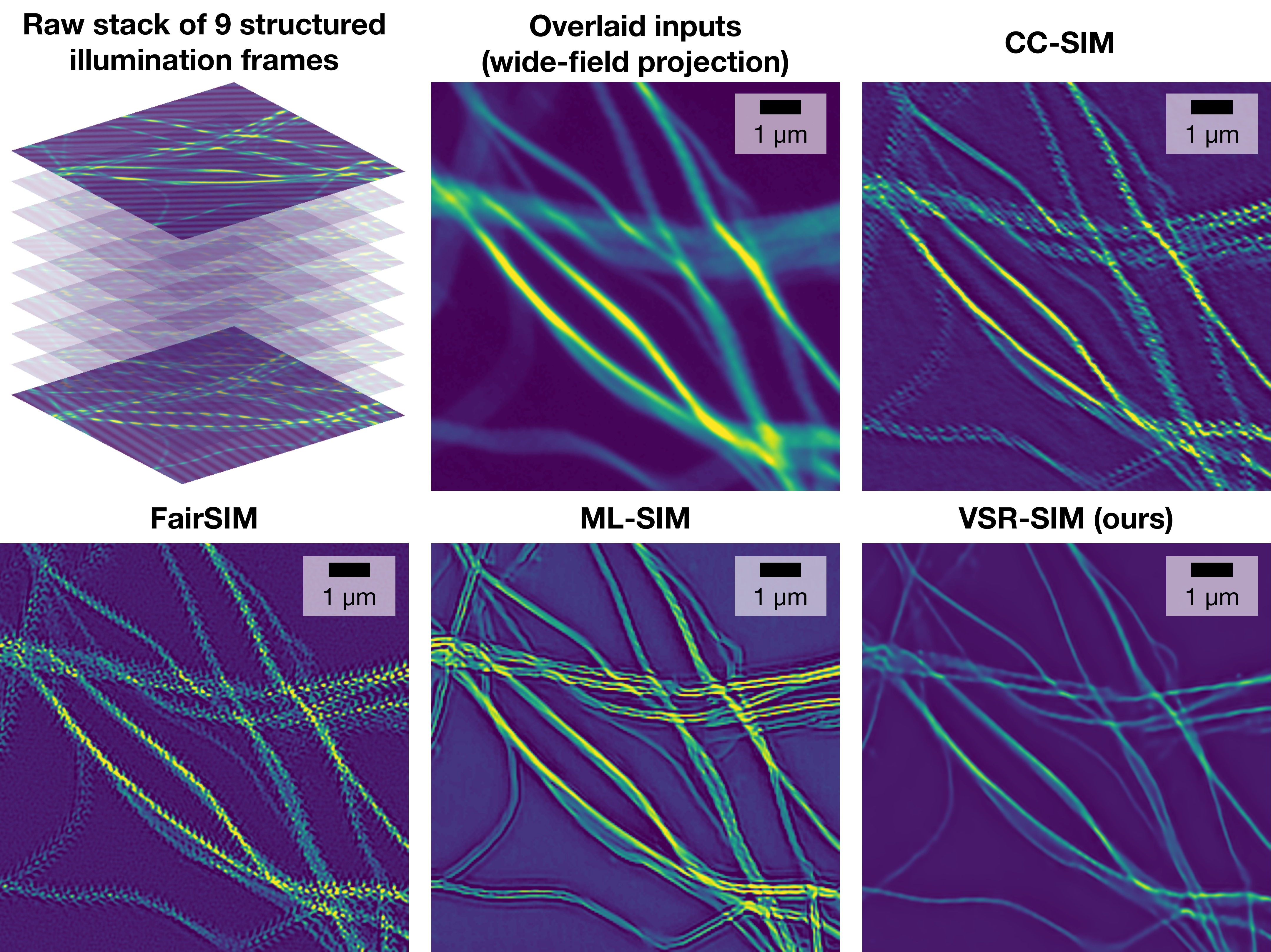}
    \caption{Structured illumination microscopy image sequences of dynamic
        samples give rise to motion artefacts when previous reconstruction
        methods are used such as cross-correlation SIM (CC-SIM) \cite{Wicker2013a},
        FairSIM \cite{Muller2016} and ML-SIM \cite{christensen2021ml}. The input
    image stack is experimental data of imaged microtubules. }
	\label{fig:expsim-comparison}
\end{figure}

The collection of SIM images corresponding to the sequence of illumination
patterns, typically a stack of 9 frames, is then used to reconstruct a
super-resolved image. Since SIM image stacks can be recorded at high frame
rates, it is possible to image highly dynamic phenomena sequentially
\cite{holcman2018single,planchon2011rapid}. However, the reconstruction methods that
are most widely used do not make use of the temporal dimension of the acquired
data \cite{Muller2016,Wicker2013a,Lal2016,huang2018fast}, because the standard
semi-analytical Fourier formalism for reconstruction assumes a static sample.  Hence, motion of the
sample between acquired frames manifests as motion blur and reconstruction
artefacts -- see \cref{fig:expsim-comparison}.

\begin{figure}[b]
	\centering
  \includegraphics[width=0.48\textwidth]{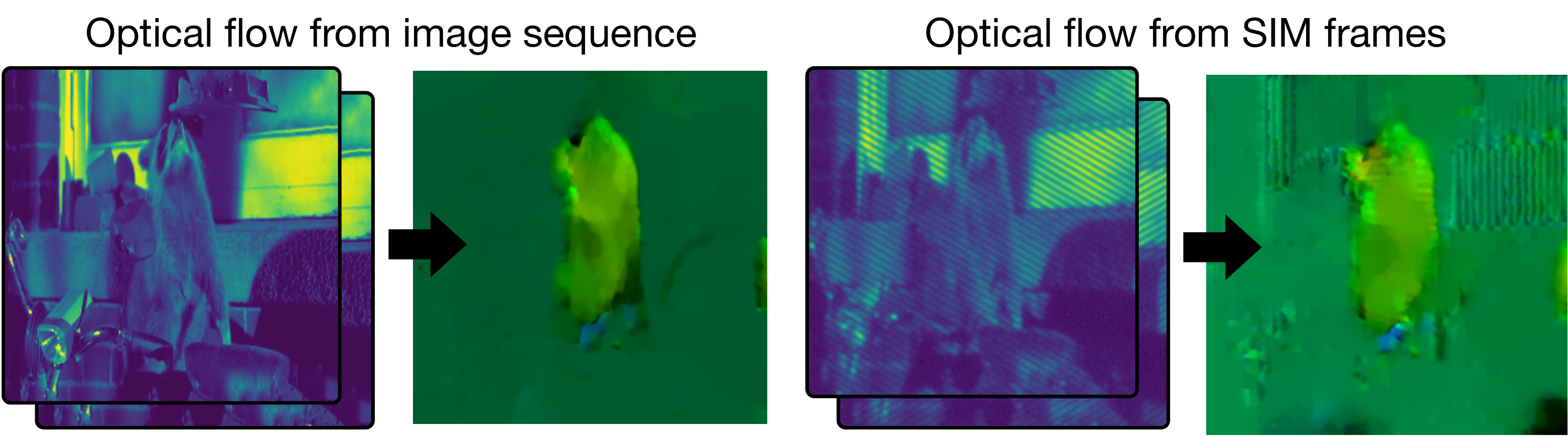}
	\caption{Optical flow computed from video-rate SIM data leads to artefacts
    that are problematic for standard motion compensation techniques.}
	\label{fig:optFlowProblem}
\end{figure}

Deep learning offers an effective way to achieve motion compensation for video
super-resolution (VSR). Recent studies demonstrate reconstruction of SIM
images using neural networks \cite{christensen2020ml,jin2020deep,ling2020fast},
offering advantages such as improved speed and robustness to noise, but none of
these reconstruction methods make use of the temporal dimension of the live-cell
data. To obtain a spatio-temporal reconstruction method for SIM, we identify the
following two problems that need to be overcome: (a) ground truth data for supervised
learning will inherit motion blur if the targets are obtained from traditional
reconstruction methods; (b) regular motion estimation methods do not work
accurately on SIM data. Machine learning implementations for SIM reconstruction
generally use as ground truth data a collection of carefully performed
reconstructions from traditional methods, which relies on an analytical
framework that assumes static samples, thus causing motion artefacts to manifest
in the training data. As for (b), a common way to incorporate high-level
reasoning about motion and occlusion in a model is bidirectional optical flow.
However, such algorithms are not directly suited for SIM imaging, because the
illumination patterns in the raw data prevent accurate calculation of motion --
the varying patterns tend to be confused with motion of the subject as
illustrated on \cref{fig:optFlowProblem}.

We propose a method to address these two problems by building upon recent
advances in using neural networks for SIM reconstruction and video
super-resolution. We generalise the approach to supervised learning proposed in
our previous method ML-SIM \cite{christensen2021ml}, in which SIM image formation is
modelled to obtain synthetic training data. Instead of simulating SIM image data
using static images, we use video sequences instead, which facilitates the
learning of motion compensation. Instead of simulating SIM image data using
static images, we use video sequences instead, such that the training
data can facilitate the learning of motion compensation. To address (b), we
propose a 3D transformer network architecture that solely relies on attention
mechanism rather than optical flow to handle subject motion.  Our contributions
are three-fold:

\begin{itemize}

    \item We demonstrate a new approach to synthesising training data
for machine learning models to learn spatio-temporal SIM reconstruction, in
which SIM image formation is simulated using video data sequences as inputs.
This enables models to be optimised for
highly dynamic sequential live-cell SIM data.
\item  We propose a video super-resolution transformer architecture that uses
	shifted windows with 3-dimensional patches to capture the spatio-temporal
	correlations in live-cell SIM data with windowed multi-head attention. We introduce
	residual connections between transformer blocks with channel attention as an additional attention
	mechanism.
	 \item We showcase a unique application of our method, rolling SIM imaging, where a moving window of frames is used for reconstruction.
	Our reconstruction method lends itself particularly well to rolling SIM
	imaging because
	it can be recast as a video super-resolution problem, where the reconstruction
	of each SIM stack uses SIM frames from the previous and subsequent SIM
	stack acquisition. This increases the temporal resolution of
	SIM imaging by a factor of 9, enabling dynamic processes in biomedical
	research to be resolved without the motion artefacts that plague previous methods.
	\end{itemize}

An online, ready-to-use and interactive implementation can be found at
\href{http://vsr-sim.github.io}{http://vsr-sim.github.io}. Source code, datasets and trained models are provided at
\href{https://github.com/charlesnchr/vsr-sim}{http://github.com/charlesnchr/vsr-sim}.

\section{Related work}

\paragraph{Optical super-resolution microscopy.}
Optical super-resolution microscopy techniques have emerged over the last three
decades to now form an essential tool for medical imaging. Several semi-analytical methods have
been proposed for SIM reconstruction
\cite{Gustafsson2000,ball2015simcheck,Wicker2013,Wicker2013a,Lal2016,Muller2016},
e.g. FairSIM and OpenSIM. These methods rely on Fourier
transformations, Wiener filters and iterative deconvolution, which can induce honeycomb and ringing
artefacts, especially when noise and motion blur are significant
\cite{Demmerle2017}. Multiple machine learning implementations for SIM
reconstruction have been proposed in the past year
\cite{jin2020deep,ling2020fast,christensen2021ml} based on
convolutional neural networks that take in SIM stacks and output super-resolved
images. Two such examples are U-Net-SIM \cite{jin2020deep} and ML-SIM
\cite{christensen2021ml} using  U-Net \cite{Ronneberger2015} and RCAN
\cite{Zhang2018a} backbones, respectively. These methods offer reconstruction
with fewer frames, higher processing speed and increased robustness to
noise compared to Fourier methods. None of these studies considered fast-moving
samples. In \cite{Huang2018}, however, SIM is applied to image highly dynamic samples
using a semi-analytical reconstruction method. This is achieved using rolling SIM
imaging, as further explored in \cref{sec:rollingShutter}, with a very short
exposure time, such that motion artefacts can be minimised. This can lead to
high frame rates, but at a significant loss of image quality, i.e. low
signal-to-noise ratio from which spatial resolution decreases. This trade-off
between temporal and spatial resolution is prevalent in the field because none
of the existing reconstruction methods for SIM exploits the spatio-temporal
nature of live-cell data. Applications of existing methods may only reduce
motion artefacts via this trade-off, whereas the capability to perform motion
compensation during reconstruction would handle these artefacts directly while
maintaining image quality.

\paragraph{Image and video super-resolution.}
Methods using convolutional neural networks as a backbone have long been
state-of-the-art for image and video super-resolution (SR). Dong \etal pioneered the
pursuit of learning-based methods for image SR by achieving superior performance
to traditional methods using a	CNN with only three layers \cite{Dong2016b}. A
similar network for VSR was proposed by Kappeler et al. \cite{Kappeler2016}.
With the emergence of residual networks \cite{He2015}, it became possible to build deeper
networks. Ledig \etal repurposed ResNet for SR with the network SRResNet
\cite{Ledig2016}. An attention mechanism was introduced by Zhang \etal
\cite{Zhang2018d} with residual channel attention network (RCAN) becoming a new
state-of-the-art method. More recently, multi-head attention has been introduced for SR
using transformer-based architectures with IPT \cite{Chen2020a} and SwinIR
\cite{Liang2021}.

For VSR, the spatio-temporal correlations between input frames are essential to
model for optimal performance. Most VSR methods use
frame alignment enabled by motion estimation and compensation \cite{Liu2020}. For motion estimation,
a popular approach is using optical flow \cite{horn1981determining}.
A state-of-the-art VSR method that uses optical flow is RBPN
\cite{Haris2019}, which is based on a recurrent CNN architecture. Recently,
the method BurstSR \cite{Bhat2021} was proposed for SR reconstruction of images
 taken in quick succession with a handheld camera. The problem is similar in
 principle to SIM reconstruction, but the method is not directly applicable as it is based on optical
 flow for alignment.
Methods that do not use optical flow tend to rely on 3D convolutional networks
 \cite{jo2018deep,Liu2021a}. However, Choi \etal demonstrated that channel attention as a
 sole mechanism is a strong baseline for motion compensation in the related problem of video
 interpolation \cite{Choi2020}.

 \paragraph{Vision transformer.} With the advent of Vision Transformer (ViT)
 \cite{Dosovitskiy2020}, transformer networks are beginning to replace CNNs for
 low-level computer vision tasks. ViT introduced multi-head self-attention
 (MSA) for image input, which proves to be a very flexible mechanism for vision,
 but does require a substantial number of trainable parameters compared with
 equivalently performing CNNs. Liu \etal demonstrated that using a hierarchy of
 shifted window MSA modules, their proposed transformer architecture, Swin, can incorporate the large receptive field of ViT, while
 having the same efficient inductive bias that CNNs offer \cite{Liu2021}.
 Variations of the Swin transformer have become state-of-the-art in image
 restoration, SwinIR \cite{Liang2021}, and video classification \cite{Liu2021a}.

%\subsection{Deconvolution}
%Deconvolution is the inverse problem of removing the blur caused by the
%point spread function of an imaging system. Deconvolution is not a direct
%alternative to SIM reconstruction, because it only offers resolution
%increase in terms of deblurring not super-resolution (sub-diffraction
%limited features). Instead it can be used in conjunction with a SIM
%reconstruction method (CITE fairsim, Florian joint-RL), which may improve
%reconstruction quality, but also introduces another level of uncertainty.
%Deconvolution as a standalone post-processing step lends itself well to
%wide-field imaging as the modality does not enable any other means of
%resolution increase. One may then consider wide-field imaging with
%deconvolution as a baseline that any SIM implementation must be held up
%to.

\section{Data generation}

Acquiring a real pairwise dataset for supervised learning in the context of
super-resolution microscopy is problematic. Experimentally, the ground truths
cannot be obtained, which leaves the options of using either the output from traditional
reconstruction methods as a target \cite{Jin2020,ling2020fast} or a different
optical super-resolution modality \cite{wang2019deep}. The former approach prevents the
method from generalising and improving beyond traditional methods, and the latter is
highly prone to artefacts, while not being live-cell
compatible. Therefore, we take the approach of generating a synthetic dataset using a SIM
image formation model \cite{christensen2021ml} on a video dataset, which
provides ideal ground truths and diverse training data.

\paragraph{Video dataset.}

\begin{table}
\begin{tabular}{|l|c|c|c|c|}
\hline
\multicolumn{1}{|c|}{\multirow{2}{*}{Test sets}}            &
\multicolumn{4}{c|}{Motion regime} \\\cline{2-5} {} &
\multicolumn{1}{l|}{\small Static} & \multicolumn{1}{l|}{\small Medium} &
	\multicolumn{1}{l|}{\small Fast} & \multicolumn{1}{l|}{\small Extreme} \\ \hline\hline
	Source      & \multicolumn{1}{c|}{DIV2K}  &
	\multicolumn{1}{c|}{BBC}      &
	\multicolumn{1}{c|}{REDS} &
		\multicolumn{1}{c|}{REDS}    \\
Data type  & Image                         & Video                            &
					 Video& Video                           \\
Frame skip  & -                         & No                            & No
						& Yes                           \\
 Samples \#  & 200                         & 50                            & 10                        & 10                           \\
Max flow   & 0                           & 10.2                          &
27.3                     & 46.2                        \\
Median flow & 0                           & 1.5                         & 10.4
						& 18.1                       \\ \hline
\end{tabular}
\caption{The four test sets that have been prepared for experiments using the
source datasets DIV2K \cite{Agustsson2017}, a subset of our BBC video dataset,
and REDS \cite{Nah_2019_CVPR_Workshops_REDS}. The
motion is amplified by skipping every other frame for the Extreme test set.
Motion is quantified by calculating max and median of the magnitude of optical flow between the
first and center frame in all sequences for a dataset at 512x512 resolution. }
\label{tab:datasets}
\end{table}

\begin{figure*}[t!]
  \includegraphics[width=\textwidth]{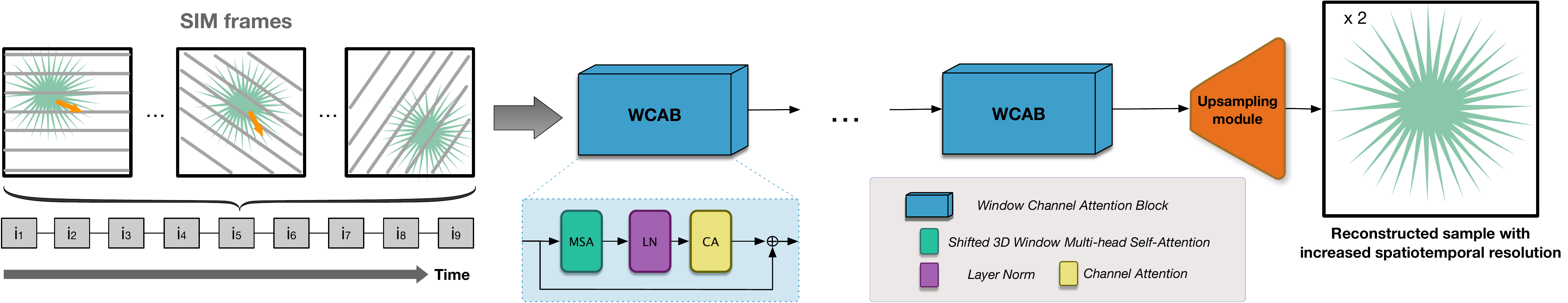}
	\caption{Architecture of the proposed windowed channel attention network.
    Skip connections are added between the attention blocks in a similar fashion
to residual networks. }
	\label{fig:arch}
\end{figure*}

Inspired by DIV2K \cite{Agustsson2017} for SISR, we built a large video dataset
focusing on diversity and high-resolution footage. Specifically, this dataset is
designed to have targets of at least 1024x1024 pixels to make the image
formation model
more consistent with typical experimental data from SIM systems, thus
facilitating model
inference performance.  Many previous VSR
datasets are limited in scope and are intended for video classification
\cite{kay2017kinetics,abu2016youtube} or more suitable for testing, e.g. REDS
\cite{Nah_2019_CVPR_Workshops_REDS}, while others only have a small subset of high-resolution, diverse data, e.g.
Vimeo90k \cite{xue2019video}. Our dataset consists of 200 hours of high-quality footage from nature
documentaries produced by the BBC. Samples are included here with permission and
video data has been obtained under the ERA License. The collection of videos is sampled to generate
100,000 image sequences, each consisting of 9 frames. A subset is reserved for
testing, for which we also use DIV2K and REDS.
The DIV2K dataset is a single
image dataset, and with the image formation model described in the
following paragraph, these still images correspond to imaging
static subjects. The REDS dataset features videos recorded with a
handheld camera with a high level of image translation from frame to frame. To
make the motion in the REDS video even more extreme, we
prepared an extra test set by sampling the videos with frame skipping, such that
only every second frame was kept. These datasets were used to prepare four test sets to assess reconstruction performance in
different motion regimes. The difficulty associated with a dataset depends on the level
of motion that its samples exhibit. We quantify this using the mean and maximum
value of optical flow magnitude averaged over all samples in the respective
datasets.  See \cref{tab:datasets} for further
specification. Further details on the dataset sampling is provided in
Supplementary.

\paragraph{Image formation model.}
SIM enables optical super-resolution by encoding structural details
corresponding to high spatial frequencies of the sample into
signals in the lower frequency domain. By unmixing the low-frequency data,
information can then be recovered that would otherwise be lost with conventional
wide-field imaging. The diffraction limit is described by the
optical transfer function (OTF), which represents the
transmittable bandwidth of spatial frequency through an imaging system. It is by shifting high spatial
frequencies into the accessible passband that super-resolution by SIM is
obtained. The OTF is the Fourier transform of the point spread function (PSF),
which is the blur kernel in direct space. Conventional wide-field SIM uses sinusoidal
 illumination patterns formed by the interference of two beams
\cite{Gustafsson2000}. The illumination pattern has an orientation and a phase
shift, which are typically varied over 3 values to ensure symmetric frequency
support, thus leading to a stack of 9 frames with different patterns. In mathematical terms, SIM
reconstruction solves the inverse problem of this
excitation and blurring operation, thus determining
the fluorescent signal that represents the sample.

The ideal OTF is generated based on a given
objective numerical aperture, pixel size and fluorescence emission
wavelength. The illumination stripe patterns are calculated from
their spatial frequency $k_0$ and phase $\phi$, \begin{align}
\label{eq-pat} I_{\theta,\phi}(x,y) = I_0 \left[ 1 - \frac{m}{2}\cos
\left( 2\pi( k_x  x + k_y y) + \phi \right) \right],
\end{align} where $k_x$, $k_y$ = $k_0 \cos \theta$, $k_0 \sin \theta$ for
a pattern orientation $\theta$ relative the horizontal axis, $\phi$
defines the phase of the pattern (i.e.\ the lateral shift in the direction
of $k_0$) and $m$ is the modulation depth, which defines the relative
strength of the super-resolution information contained in the raw images.
 The fluorescent
response of the sample can then be modelled by the multiplication of the
sample structure, $S_t(x,y)$, i.e. input image, at time $t$ and the illumination pattern
intensity $I_{\theta,\phi}(x,y)$. The final image, $D_{t,\theta,\phi}(x,y)$,
is formed after blurring by the PSF, $H(x,y)$, and addition of
white Gaussian noise, $N(x,y)$, \begin{align} \label{eq-irf}
D_{t,\theta,\phi}(x,y) = \left[ S_t(x,y) I_{\theta,\phi} \right] \otimes
H(x,y) + N(x,y), \end{align} where $\otimes$ is the convolution operation.
The set of sampled images from a sequence in the video dataset corresponds to the time
points $t \in \left[1,9\right]$.
A full SIM stack is comprised of the set $\left\{D_{t,\theta,\phi}\, \vert \, t \in
\left[1,9\right]\right\}$, where each value of $t$ is associated with a distinct
illumination pattern, i.e. a unique
permutation of $\theta$ and  $\phi$. Each consecutive 9 frames then contain a
full cycle of  illumination patterns.
In addition to Gaussian noise, added pixel-by-pixel,
a random error is added to the parameters for the stripe patterns, $k_0$,
$\theta$ and $\phi$, to approximate the inherent uncertainty in an
experimental setup for illumination pattern generation as well as forcing the
model to generalise when learning the reconstruction task from the data. Poisson noise can
further be introduced to more realistically approximate the noise sources
present in experimental data. For implementation details and specification of optical
parameters see Supplementary.

\section{Model architecture} The proposed model is inspired by the vision
transformer network \cite{Dosovitskiy2020} in particular its more efficient
shifted window  variant, Swin \cite{Liu2021},  with its extension for video
classification, Video Swin \cite{Liu2021a}, and adaption to image restoration,
SwinIR \cite{Liang2021}. Swin introduced the inductive bias to self-attention
called shifted window multi-head attention (SW-MSA), which can be compared to
the inductive bias inherent to convolutional networks.  SwinIR introduced
residual blocks to the Swin transformer to help preserve high-frequency
information for deep feature extraction. The Video Swin transformer generalised
the SW-MSA to three dimensions, such that spatio-temporal data can be included
in the local attention for the self-attention calculation. Further to this, the
success of the  channel attention mechanism in \cite{Zhang2018d} inspires the
inclusion of this other inductive bias in addition to 3D local self-attention
following the SW-MSA approach.

The inputs to the model have dimension $T\times H \times W \times C$, where $T$ is 9 for SIM
reconstruction and $C$ is 1. A shallow feature extraction module in the beginning of the network architecture
\cref{fig:arch} projects the input into a feature map, $F_0$, of $T \times H
\times W \times D$
dimension, where the embedding dimension, $D$, is a hyperparameter.
The feature map is passed through a sequence of residual blocks, denoted
Window
Channel Attention Block (WCAB)
\begin{align}
	F_i = H_{\mathrm{WCAB}}(F_{i-1}),\quad i = 11,..,n
\end{align}
Inside each WCAB is a sequence of Swin Transformer Layers (STLs), in which
multi-head self-attention is calculated using local attention with shifted window
mechanism. Inputs to STL layer is partitioned into
$\frac{T}{P}\times \frac{HW}{M^2} $ 3D tokens of $P\times M^2\times D$ dimension.
For a local window feature, $x\in \mathbb{R}^{P\times M^2\times D}$, query, key
and value matrices, $\left\{Q,K,V\right\} \in \mathbb{R}^{PM^2\times  D}$, are
computed by multiplication with projection matrices following the original
formulation of transformers \cite{Vaswani2017}. Attention is then computed as
\begin{align}
	\mathrm{Attention}(Q,K,V) = \mathrm{SoftMax}(QK^T/ \sqrt{d} + B)V,
\end{align}
where $B\in\mathbb{R}^{P^2\times M^2\times M^2}$ is a relative positional bias
found to lead to significant improvements in
\cite{Liu2021}.
 STLs are joined in a way similar to the residual blocks, although the use of
SW-MSA is alternated with a version without shifted windows, W-MSA, ensuring
that attention is computed across window boundaries, which would not have been
the case without SW-MSA.

After the final STL, the $m$-th layer, in a WCAB, a transposed 3-dimensional convolutional layer is
used to project the 3D tokens back into a $T\times H\times W\times D$ feature map, $F_{i,m}$. A
channel attention module is then used on $F_{i,m}$ to determine the dependencies
between channels following the calculation of the channel attention statistic
\cite{Zhang2018d}. The mechanism works by using global adaptive average pooling
to reduce the feature map to a vector which after passing through a
2D convolutional layer becomes  weights that are multiplied back onto $F_{i,m}$,
such that channels are adaptively weighed. A residual is then obtained by adding a
skip connection from the beginning of
the $i$-th WCAB to prevent loss of information, i.e. low-frequency information,
and the vanishing gradient problem.
A fusion layer combines the temporal dimension and the channel dimensions.
For the final upsampling module, we use the sub-pixel convolutional filter
\cite{Shi2016} to expand the image dimensions by aggregating the fused feature
maps. The implementation will be made available on Github and is provided in
the Supplementary.

\begin{figure}[h!]
		\includegraphics[width=0.5\textwidth]{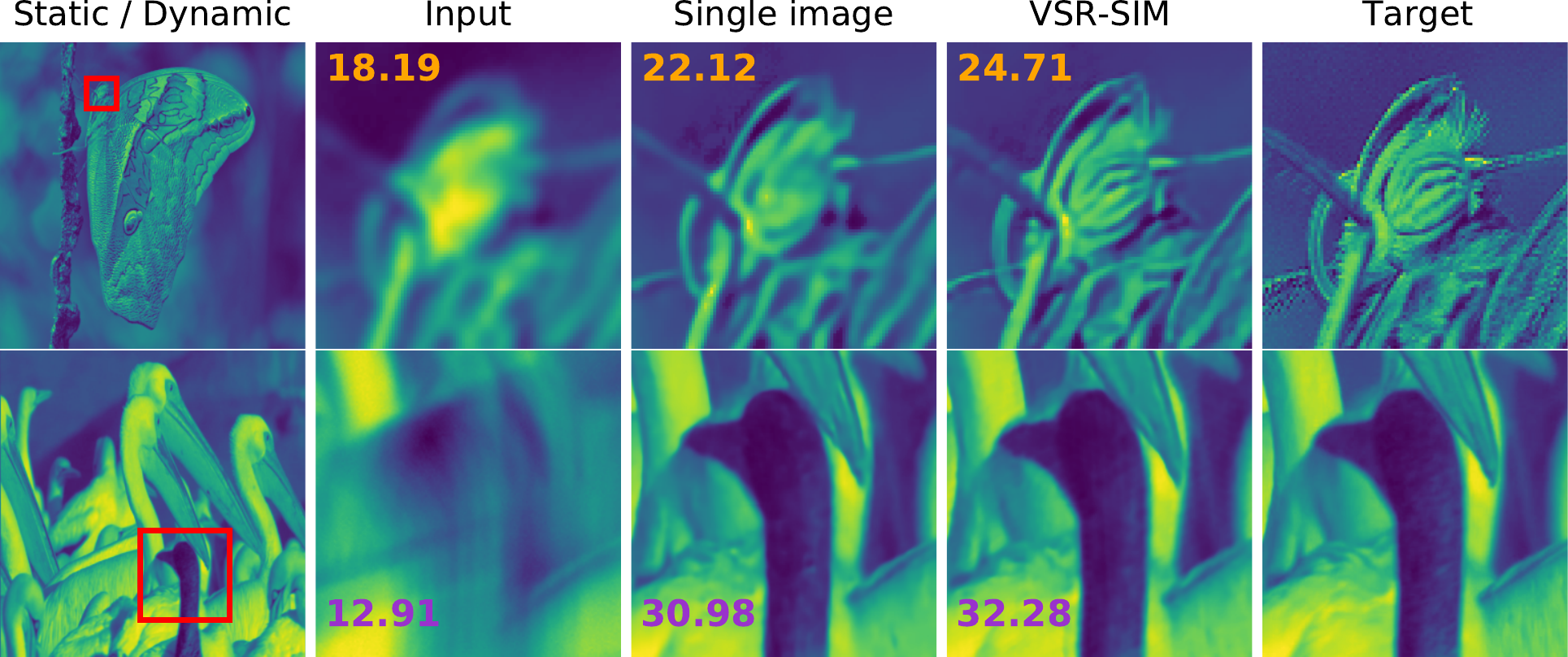}
	\caption{For static subjects (top row) the method defaults to standard SIM
	reconstruction, which offers significant improvements over a deconvolution
	baseline trained with the same architecture. For dynamic input data (bottom
	row), the advantage of SIM diminishes depending on the level of motion, but,
	importantly, VSR-SIM does not generate motion artefacts in this setting.    }
	\label{fig:static_comparison}
\end{figure}

\section{Experiments}

\paragraph{Implementation details.} All models described in the following were
trained using the Adam optimiser and a mean squared error loss function with a
learning rate of 1e-4 that is halved every
100,000 iterations. A total of 500,000 iterations were made, which equals 5
epochs of the BBC training dataset. A set of 4 Nvidia A100 GPUs was used with a
batch size per GPU of 4. Training samples were randomly cropped to 128x128 inputs
and 256x256 targets, while inference was performed with 512x512 inputs resulting in 1024x1024
outputs. For VSR-SIM, the WCAB number, STL number, window size, embedding size
$D$ and attention head number are set to 6, 6, 8, 96 and 6, respectively. The
hyperparameters of the other tested architectures follow original
implementations and are further specified in Supplementary.

\begin{table}[h]
	\centering
\begin{tabular}{|l|c|c|}
\hline
\multicolumn{1}{|l|}{\multirow{2}{*}{Reconstruction method}}      &
\multicolumn{2}{|c|}{Test set (PSNR)}  \\\cline{2-3}
\multicolumn{1}{|c|}{}           & \multicolumn{1}{l|}{Static} &
\multicolumn{1}{l|}{Medium} \\ \hline\hline
Wide-field baseline & 22.79                            & 17.31                             \\
CC-SIM \cite{Wicker2013a}                & 27.99                            & 16.98                             \\
OpenSIM \cite{Lal2016}               & 28.34                            & 14.04                             \\
FairSIM \cite{Muller2016}               & 28.54                            & 15.34                             \\
ML-SIM \cite{christensen2021ml}                & 32.30                             & 18.41                             \\
VSR-SIM (ours)               & \textbf{34.74}                            &
\textbf{30.15}                             \\ \hline
\end{tabular}
\caption{Synthetic test sets were evaluated with four existing SIM reconstruction
methods and VSR-SIM. The static test set was generated using still images from
DIV2K \cite{Agustsson2017} and
the dynamic test set was generated using image sequences sampled from the BBC video dataset.
At high levels of motion, other SIM reconstruction methods fail, but VSR-SIM can
maintain a high reconstruction quality for the dynamic test set.}
\label{tab:expsim}
\end{table}

\begin{figure}[h]
	\centering
  \includegraphics[width=0.48\textwidth]{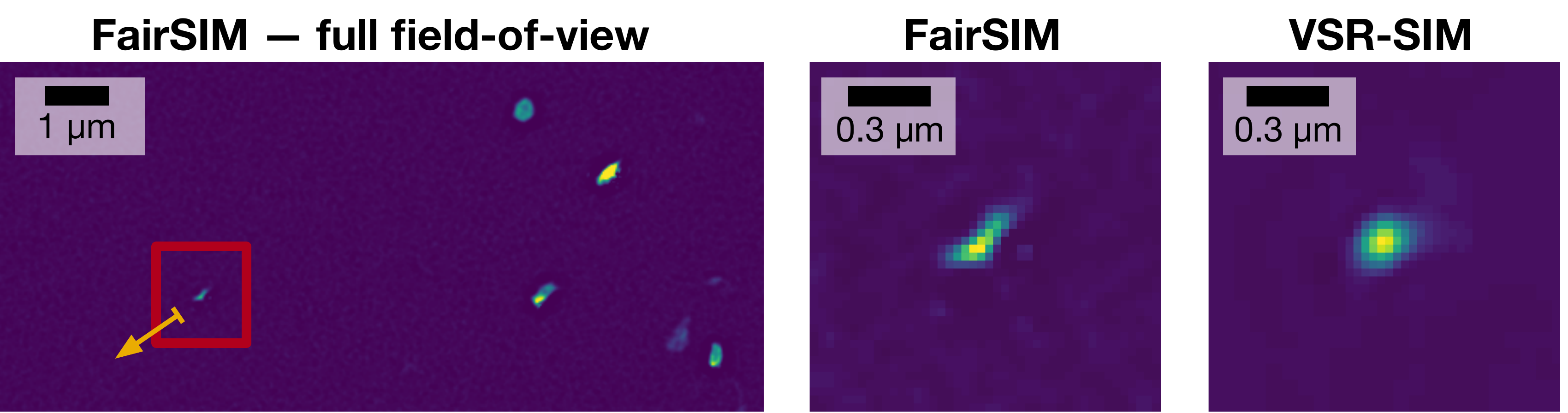}
	\caption{Lysosome, a spherical vesicle, moving rapidly in a sample of COS-7
        cells. FairSIM is unable to handle motion blur and reconstructs an elongated
shape, while VSR-SIM reconstructs a circular shape consistent with the known
shape of the lysosome.}
	\label{fig:temporalResolutionFig}
\end{figure}

\begin{table}
	\centering
\begin{tabular}{|l|c|c|c|c|}
\hline
\multicolumn{1}{|l|}{\multirow{2}{*}{Method}}      &
\multicolumn{4}{|c|}{Test set (PSNR)}  \\\cline{2-5}
\multicolumn{1}{|c|}{}  & \multicolumn{1}{l|}{Static} &
\multicolumn{1}{l|}{Medium} & \multicolumn{1}{l|}{Fast} &
\multicolumn{1}{l|}{Extreme} \\
\hline\hline
Bicubic${}^\dagger$ & 26.40                                & 26.35
															& 22.63                           & 21.08
															\\
SISR${}^\dagger$                  & 31.23                               & 28.08
															& 25.38                           & 22.50                                                       \\
VSR${}^\dagger$                 & 31.15                               & 28.15                                 & 25.41                           & \textbf{22.98}                                             \\
VSR-SIM                       & \textbf{34.74}                      & \textbf{30.15}                        & \textbf{26.04}                  & 22.95                                                      \\
RBPN                          & 33.16                               & 29.25                                 & 25.29                           & 21.48                                                      \\
Wide-field        & 26.24                               & 22.99                                 & 19.32                           & 18.77                                                      \\ \hline
\end{tabular}
\caption{Test of our method in different motion regimes compared with
	baseline models trained and evaluated using input without structured
	illumination. Methods denoted with ${}^\dagger$ are based on
input without illumination patterns. The SISR and VSR baseline methods use
the same architecture as VSR-SIM. The sub-diffraction limit resolution of SIM
is achievable despite significant motion in the input data, but is ultimately lost for an extreme
level of motion. RBPN that uses optical flow for motion estimation was
not found to perform comparably, suggesting that optical flow is not needed. }
\label{tab:motionregimes}
\end{table}

\begin{figure*}[t]
	\centering
  \includegraphics[width=1\textwidth]{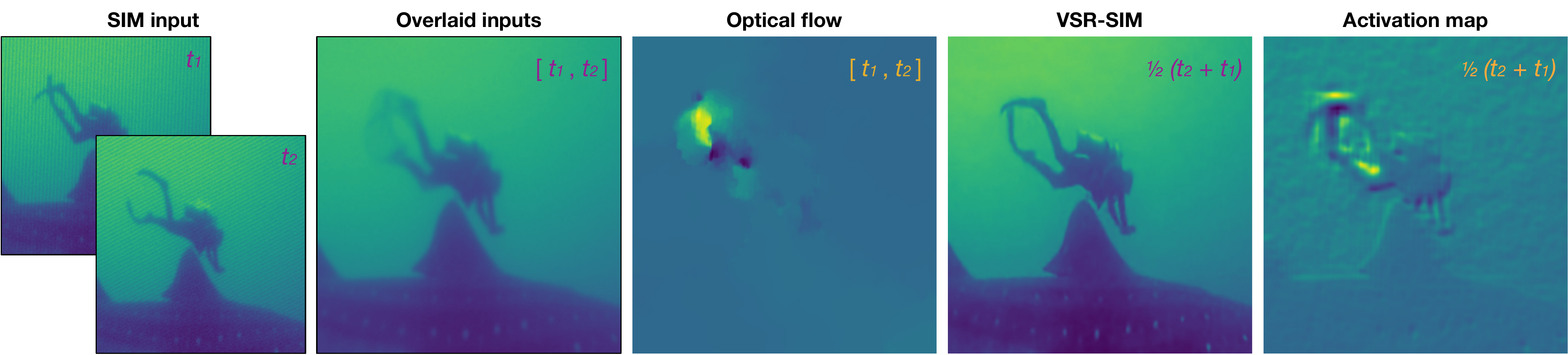}
	\caption{Self-attention appears to emphasise the regions, in which motion occurs. The activations from the final attention heads are found to be well correlated with intensity maps of optical flow.}
	\label{fig:optFlow_vs_activations}
\end{figure*}

\subsection{Comparison with state-of-the-art}

\paragraph{SIM reconstruction methods.}

The Static and Medium test sets, see \cref{tab:datasets} for details,  were evaluated with our method, VSR-SIM, and four
existing SIM methods: CC-SIM \cite{Wicker2013a}, OpenSIM \cite{Lal2016}, FairSIM
\cite{Muller2016} and ML-SIM \cite{christensen2021ml}. The results are listed in
\cref{tab:expsim} based on peak signal-to-noise ratio (PSNR). For the Static test set the difference in reconstruction
quality is relatively even, but for the Medium test set, most previous methods fail
to surpass the diffraction-limited wide-field baseline. This is due to motion artefacts
and inaccurate numerical optimisation (e.g. parameter estimation using
peak finding in the case of FairSIM) becoming substantial. An example
illustrating motion artefacts in reconstruction output
for an input sample with significant motion is shown in \cref{fig:expsim-comparison}.

We tested the spatio-temporal resolution of reconstruction on a real sample by imaging fast-moving
lysosomes within the endoplasmic reticulum (ER) in COS-7 cells
\cite{lu2020structure}. We use the SiR-lysosome fluorophore with an excitation wavelength of  652 nm. Given the same raw data, we observe
differences in the shape of the lysosome following reconstructing with FairSIM and
VSR-SIM, see \cref{fig:temporalResolutionFig}. FairSIM produces an elongated
shape suggesting that motion blur is reconstructed into features, which is
further supported by the simulated test in \cref{sec:speedLimit}.

%\begin{table*}[h]
%\begin{center}
%\begin{tabular}{c|c|c|c|c}
%\hline
%\# & Method & Training Data & Initialization & Test Accuracy\\
%\hline\hline
%1 & AlexNet & $\mathcal{D}_c$ & random & 64.54\% \\
%2 & AlexNet & $\mathcal{D}_c$ & ImageNet pretrained & 72.63\% \\
%3 & AlexNet & upsampled $\mathcal{D}_c$ and $\mathcal{D}_{\eta}$ as ground truths & random & 74.03\% \\
%4 & AlexNet & upsampled $\mathcal{D}_c$ and $\mathcal{D}_{\eta}$ as ground truths & ImaegNet pretrained & 75.13\% \\
%5 & AlexNet & upsampled $\mathcal{D}_c$ and $\mathcal{D}_{\eta}$ as ground truths & model \#2 & 75.30\% \\
%6 & Pseudo-Label\cite{lee2013pseudo} & upsampled $\mathcal{D}_c$ and $\mathcal{D}_{\eta}$ as unlabeled & model \#2 & 73.04\% \\
%7 & Bottom-Up~\cite{sukhbaatar2014learning} & upsampled $\mathcal{D}_c$ and $\mathcal{D}_{\eta}$ & model \#2 & 76.22\% \\
%8 & Ours & upsampled $\mathcal{D}_c$ and $\mathcal{D}_{\eta}$ & model \#2 & \textbf{78.24\%} \\
%\hline
%\end{tabular}
%\end{center}
%\caption{Experimental results on the clothing classification dataset. $\mathcal{D}_c$ contains $47,570$ clean labels while $\mathcal{D}_{\eta}$ contains $10^6$ noisy labels.}
%\label{tab:exp}
%\end{table*}

%\begin{table}
	%\centering
	%\begin{tabular}{lcc}
		%\toprule
		%& \multicolumn{2}{c}{Data} \\ \cmidrule(lr){2-3}
		%Name & Column 1 & Another column \\
		%\midrule
		%Some data & 10 & 95 \\
		%Other data & 30 & 49 \\
		%\addlinespace
		%Different stuff & 99 & 12 \\
		%\bottomrule
	%\end{tabular}
	%\caption{My caption.}
	%\label{tab-label}
%\end{table}

\begin{table}[h]
	\centering
\begin{tabular}{|c|c|c|c|}
\hline
CA         & SW-MSA        & 3D window
					 & \multicolumn{1}{l|}{Score (PSNR)} \\
													\hline\hline
\checkmark &                           &                           & 29.06                                  \\ \hline
                          & \checkmark &                           & 29.48                                  \\ \hline
                          & \checkmark & \checkmark & 29.10                                   \\ \hline
\checkmark & \checkmark &                           & 30.01                                  \\ \hline
\checkmark & \checkmark & \checkmark & \textbf{30.15}                         \\ \hline
\end{tabular}
\caption{Ablation study on the inclusion of different attention mechanisms. CA is
channel attention \cite{Zhang2018d}, SW-MSA \cite{Liu2021} and 3D window refers
to 3D window attention for spatio-temporal data \cite{Liu2021a}. The scores are
based on evaluations on the Medium test set.}
\label{tab:mechanisms}
\end{table}

\subsection{Ablation study}
\label{sec:ablation}

\begin{figure*}[h!]
	\centering
		\includegraphics[width=1\textwidth]{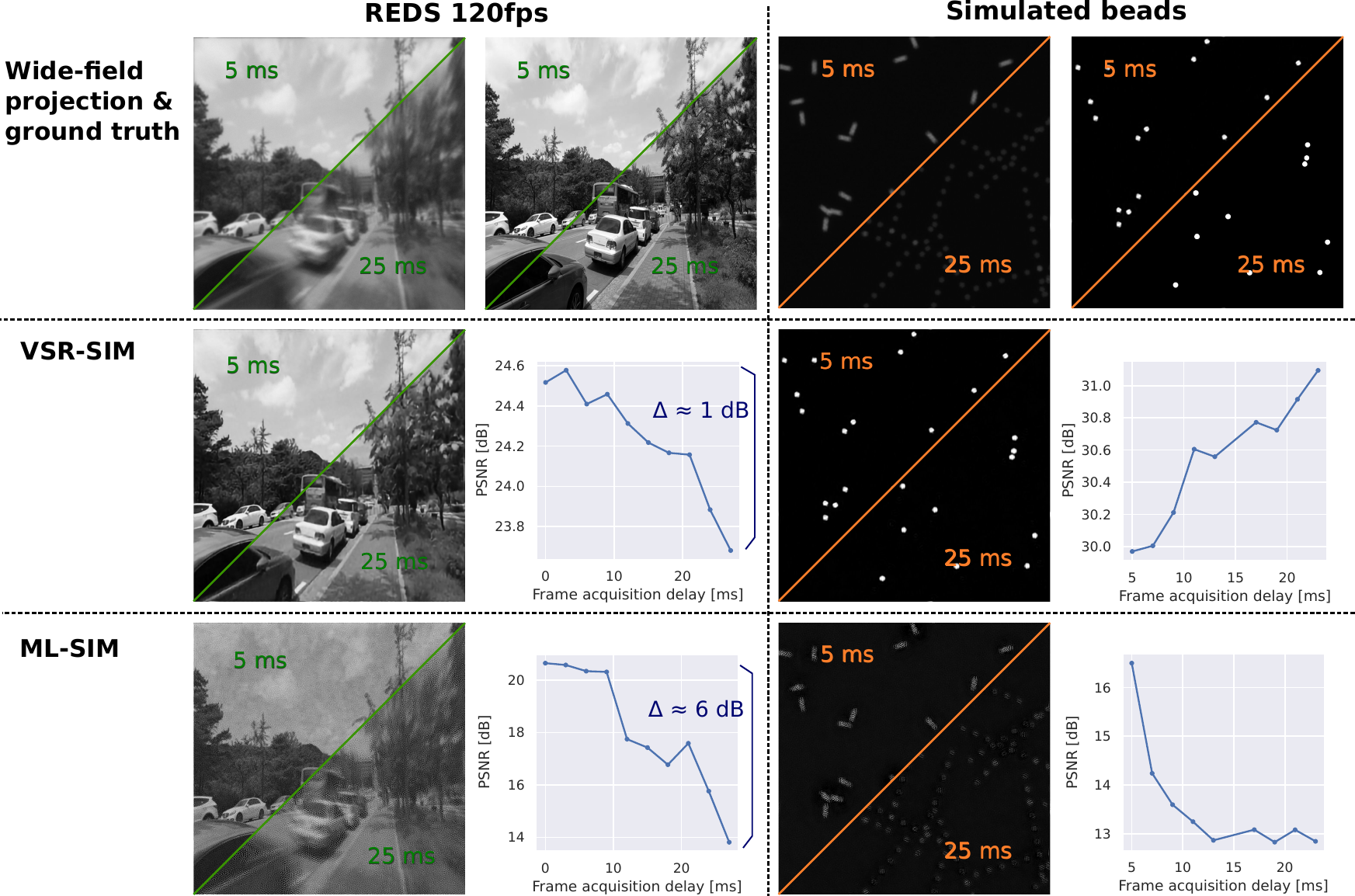}
	\caption{Reconstruction performance for VSR-SIM does not collapse for
    inputs that exhibit significant levels of motion. Given the same inputs
    sequences, the
motion can be controlled via a set delay between frames. This is done with frame
skipping for a high frame rate video sequence, REDS 120fps
\cite{Nah_2019_CVPR_Workshops_REDS}, and sequences of simulated beads. }
	\label{fig:speedAnalysis}
\end{figure*}

\paragraph{No structured illumination patterns.} An important baseline for
SIM reconstruction is deconvolution, which in this context is considered a
deblurring operation that does not
need patterned
illumination. A single-image deconvolution
method is useful for wide-field imaging to counter the effect of the PSF and
noise sources, but it cannot provide optical SR. We trained a model with the
same architecture as VSR-SIM using an equivalent dataset without illumination
patterns to synthesise wide-field images. On \cref{fig:static_comparison}, example output can be seen
showing the SISR baseline model versus VSR-SIM that takes SIM input. In the first
input sample, the subject is static, and the quality difference of the outputs
is significant. For more dynamic subjects, the difficulty of the SIM
reconstruction problem increases, and the difference to the SISR baseline is
smaller. We explored this further by testing models on the four test sets shown
in \cref{tab:datasets}. The four test sets are evaluated with a deconvolution SISR
baseline, a deconvolution VSR baseline, a state-of-the-art VSR method RBPN \cite{Haris2019}
and our method VSR-SIM. The two baseline models are based on the VSR-SIM
architecture but trained and tested without illumination patterns, while RBPN
and VSR-SIM are trained with SIM inputs. Only the center frame in a sequence
corresponding to the target is input to the SISR model, whereas the VSR model
works on the full image sequence. The test results in \cref{tab:motionregimes}
show that VSR-SIM enables high-quality SIM reconstruction in every motion
regime. The quality of the reconstruction outputs is markedly better than for
the baselines in all but the most extreme case with frame skipping.  Hence, at
very high levels of motion, the SIM modality does not offer an advantage over
conventional imaging.  This is consistent with previous theoretical findings of
Ströhl and Kaminski
\cite{Strohl2017b}.

\paragraph{Optical flow.}
As illustrated in \cref{fig:optFlowProblem} the determination
of optical flow can be hindered by the presence of an illumination pattern. The
quantitative
impact of including optical flow is tested by training
RBPN, which uses optical flow for inputting aligned frames into a recurrent network in a
mechanism called back-projection. In \cref{tab:motionregimes} it is found that the VSR-SIM model outperforms
RBPN in different motion regimes although it does not use optical flow. This
indicates that the two
attention mechanisms of VSR-SIM are sufficient to attend to regions that exhibit
a lot of motion. This is further explored by visualising the activation maps
from the final attention heads
in the network, see \cref{fig:optFlow_vs_activations}. Comparing the two frames
for $t_1$ and $t_2$, it is clear that the motion in this sequence occurs in a
very specific region, which is picked up by the optical flow intensity
projection as
well as the activation map.

\paragraph{Attention mechanisms.} The respective importance of multi-head self-attention,
3D window attention and channel attention is investigated by training different
variants of the model on the same training dataset and testing it with our
Medium test set. The results are summarised in \cref{tab:mechanisms}. The most
significant mechanism according to these results is the multi-head self-attention,
which is implemented in a similar way to SwinIR \cite{Liang2021} when 3D window attention is
excluded.

\begin{figure}[b!]
\includegraphics[width=0.5\textwidth]{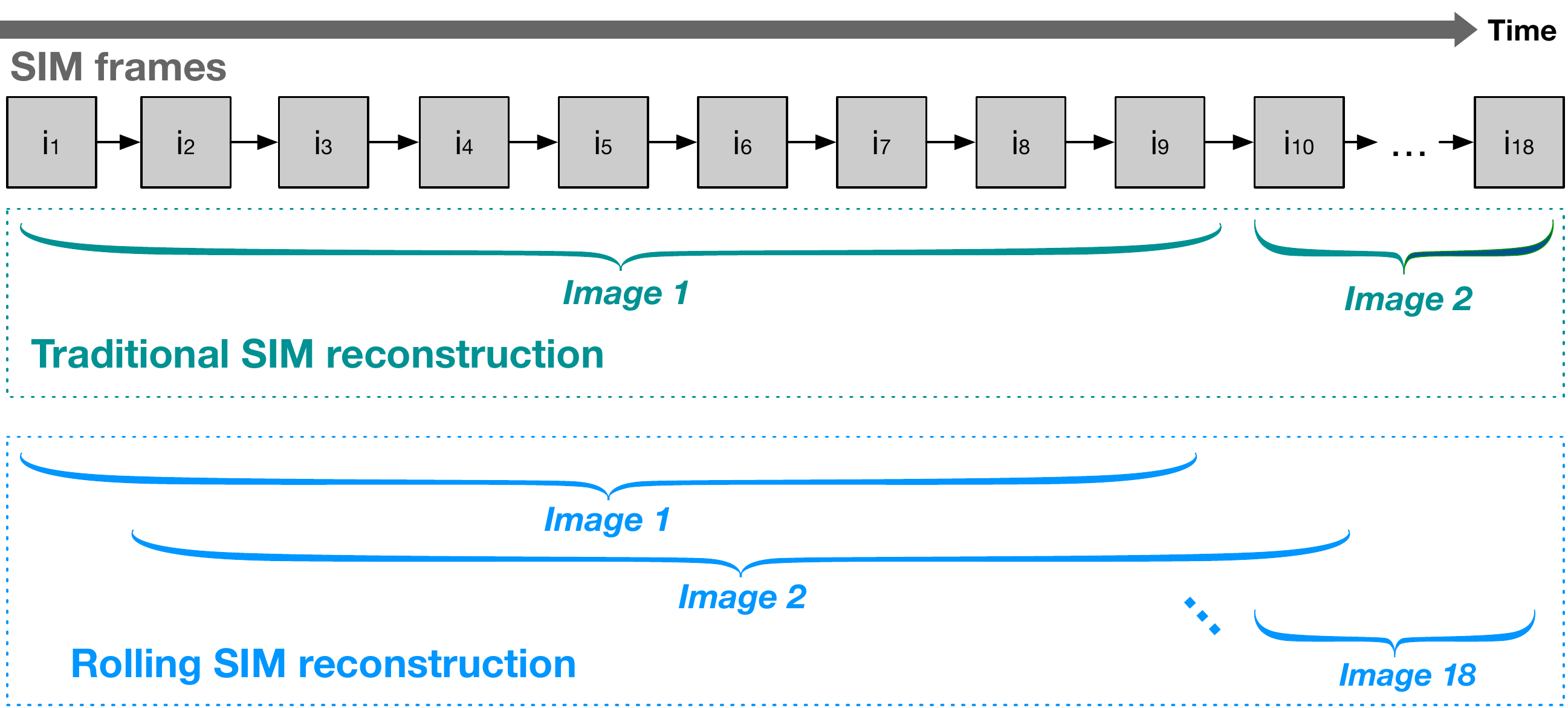}
\caption{Rolling SIM imaging scheme for structured illumination
microscopy, which is utilised in the proposed method. }
\label{fig:rolling}
\end{figure}

\begin{figure*}[h]
	\centering
		\includegraphics[width=1\textwidth]{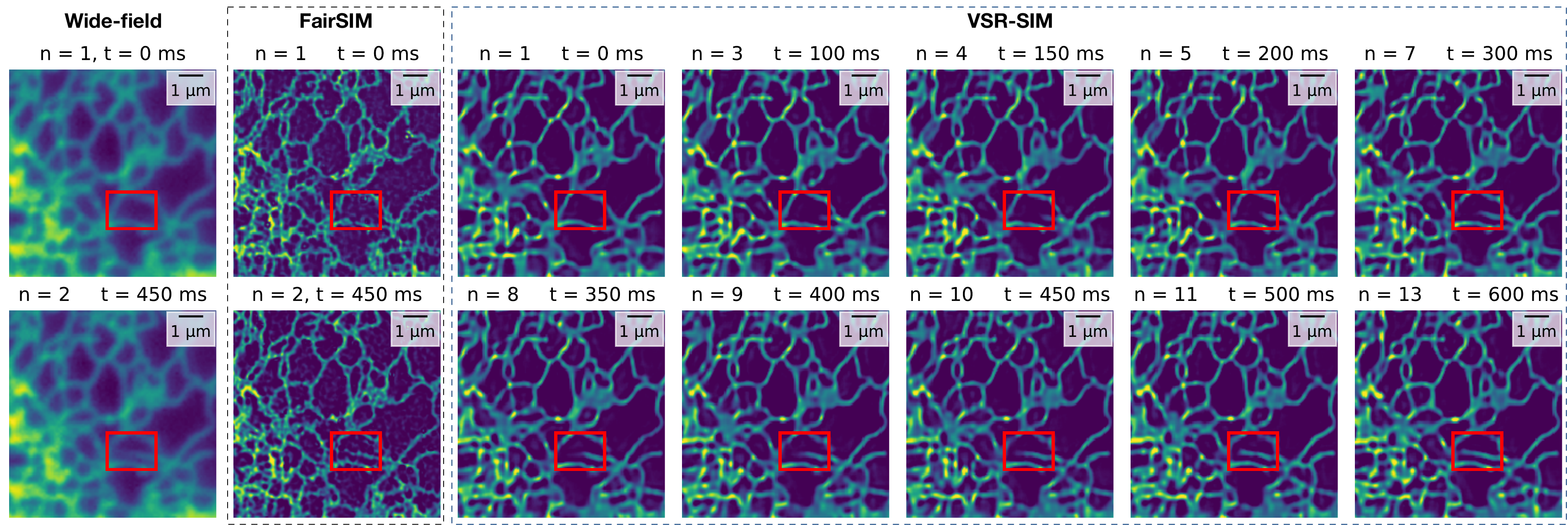}
	\caption{Our method, VSR-SIM, and the widely used method FairSIM applied to a SIM
		image
	sequence of the endoplasmic reticulum. Both methods offer significant
    improvements over wide-field imaging. The rectangle emphasises a reshaping event of
a tubule. Compared with FairSIM,  our method
achieves 9 times higher temporal resolution by enabling the rolling SIM imaging
scheme, see \cref{fig:rolling}. The spatial resolution of FairSIM is
higher, but the data contains more artefacts.  }
	\label{fig:increasedTemporal}
\end{figure*}

\subsection{Speed limit of SIM reconstruction}
\label{sec:speedLimit}

As indicated in Table \cref{tab:motionregimes}, the reconstruction
quality of VSR-SIM approaches that of a similarly trained deconvolution method,
meaning that the sub-diffraction imaging enabled by SIM becomes increasingly
difficult to achieve as the motion increases. Importantly, however, since
VSR-SIM is trained on SIM video data spanning multiple motion regimes, the case
of extreme motion does not cause the method to collapse and perform significantly worse than
the deconvolution baseline. We investigate this ability further by
reconstructing inputs that have variable
delay between frames and comparing the results to those of ML-SIM, which has no
capability to handle motion. As the input data we sample from a high frame rate video sequence from Reds
\cite{Nah_2019_CVPR_Workshops_REDS} and generate images of moving simulated beads. The
results are shown on \cref{fig:speedAnalysis}. Although the performance
decreases as
the frame delay increases, the drop is much smaller than for ML-SIM; namely 1
dB versus 6 dB over the range of 0-25 ms frame delay in case of the video
sequence from Reds. In case of
the simulated beads, the performance does not decrease. This indicates that
VSR-SIM is able to ignore the adjacent frames in a SIM stack if the
motion is high enough, which presumably becomes easier for the model to do as
the spatial separation between the beads increases.

\subsection{Rolling SIM algorithm}
\label{sec:rollingShutter}
When performing SIM reconstruction with conventional methods, the order of
illumination patterns in a stack has to be consistent across stacks.  To increase the temporal resolution of SIM, one can use frames that
belong to adjacent stacks, thus having a rolling window for which frames are
included in the current stack, which reduces the number of frames to be acquired per
individual stack. This scheme for SIM imaging is illustrated in
\cref{fig:rolling}. In the scheme depicted here, the rolling window is shifting
by
a single frame at a time, therefore increasing the temporal resolution by a
factor of 9. To reconstruct SIM frames according to a rolling window, the
reconstruction method must be able to handle inputs with varying order of
illumination patterns. We address this by shuffling illumination patterns for
every training sample that is generated for the training data. The shuffling is
without replacement such that a complete cycle is always present in an input. This
approach forces the model to learn to handle arbitrary orderings facilitating
the rolling SIM scheme. Combined with the motion compensating reconstruction
method that can work at motion regimes that traditionally would be
unmanageable, imaging at high speed with high granularity becomes
possible. This capability lends itself well to applications with fast-moving
samples that exhibit intricate movement behaviour. The scheme can similarly be
applied for long-term imaging by utilising the higher photon efficiency coming
with acquiring only a single frame per reconstructed output.

\paragraph{Improving temporal resolution.} To demonstrate our
model applied to the rolling SIM scheme, we performed an experiment imaging
endoplasmic reticulum in COS-7 cells, labelled with the sec61-mApple and imaged with an excitation
wavelength of 561 nm. The FairSIM
reconstruction method \cite{Muller2016} is used as a baseline as it is widely
used in the microscopy community \cite{samanta2021overview}. The endoplasmic
reticulum is the largest membrane structure inside the cell and displays
drastic reshaping with constant tubule elongation, retraction and junction
formation as shown on \cref{fig:increasedTemporal}. This dynamic reshaping is
important to regulate the morphology and function of ER inside the cell. Compromised
reshaping dynamics of ER is associated with a variety of diseases, including Alzheimer’s
disease \cite{yang2007reticulons}, which makes it important to record,
measure and understand these dynamics.  On
\cref{fig:increasedTemporal} an occurrence of reshaping can be seen in the area
marked by the rectangle over a sequence of 20 frames each acquired with a 50 ms
exposure time. Using FairSIM for the reconstruction provides only two
super-resolved SIM images, rendering the reshaping event very abrupt and less
noticeable. Using VSR-SIM with the rolling SIM scheme, the raw sequence leads to
19 reconstructed outputs, of which 12 are included showing a significantly more
granular process. FairSIM, however, is seen to recover more high-frequency
information in its two outputs indicating that it achieves a higher spatial
resolution, although at the expense of more artefacts.

\section{Conclusion} We have proposed a new transformer architecture that combines channel attention with multi-head
self-attention computed using shifted 3D windows. This architecture is
shown to excel at the SIM reconstruction task for dynamic inputs.
A demonstration of using the method for a use case in medical research is
made with implementation of rolling SIM imaging, in which a moving window of
SIM frames is used for reconstruction providing a temporal resolution that is 9
times higher, while still providing comparable spatial resolution well beyond
the diffraction limit.
Our method can be used for any SIM imaging system as it is purely trained on synthetic data
using our image formation model that can be easily adapted to different SIM configurations.

\section*{Acknowledgments}

CFK acknowledges funding from the UK Engineering and Physical Sciences Research
Council, EPSRC (grants EP/L015889/1 and EP/H018301/1), the Wellcome Trust
(grants 3-3249/Z/16/Z and 089703/Z/09/Z) and the UK Medical Research Council,
MRC (grants MR/K015850/1 and MR/K02292X/1), MedImmune, and Infinitus (China)
Ltd. The Research Computing Services of the University of Cambridge enabled the
high-performance computations in this work.

\section*{Disclosures}

\medskip

\noindent\textbf{Disclosures.} The authors declare no conflicts of interest.
%------------------------------------------------------------------------

%%%%%%%%% REFERENCES
\bibliography{library}
\bibliographyfullrefs{library}

% \clearpage

% \includepdf[pages=1-,pagecommand={\thispagestyle{empty}}]{supplementary/master.pdf}

\end{document}